\documentclass[aps,prb,unsortedaddress, %superscriptaddress,
               showpacs,footinbib,amsfonts,amssymb,amsmath,nobibnotes,
               reprint,
              ]{revtex4-2}

\usepackage[utf8]{inputenc}
\usepackage{amsmath,amsfonts,amsthm} 
\usepackage{amssymb}
\usepackage[svgnames]{xcolor}

\usepackage{fix-cm}
\usepackage{graphicx}
\usepackage{listings}
\usepackage{mwe}
\usepackage{booktabs} % Horizontal rules in tables
\usepackage{float} % Required for tables and figures in the multi-column environment - they need to be placed in specific locations with the [H] (e.g. \begin{table}[H])

\usepackage[colorlinks=true,linkcolor=blue,citecolor=blue,urlcolor=blue]{hyperref}
\usepackage[hyphenbreaks]{breakurl}

\graphicspath{{img/}}

\usepackage[version=4]{mhchem}

\newcommand{\uqtr}{%
  D\'epartement de Chimie, Biochimie et Physique, Institut de recherche sur l'hydrog\`ene, Universit\'e du Qu\'ebec \`a Trois-Rivi\`eres, Trois-Rivi\`eres, Canada
  }

\begin{document}

\title{Electronic transport in titanium carbide MXenes from first principles}

\author{Nesrine Boussadoune}
\author{Olivier Nadeau}
\author{Gabriel Antonius}
\affiliation{\uqtr}

\begin{abstract}

We compute from first principles the electronic, vibrational, and transport properties of four known MXenes: \ce{Ti_{3}C_{2}}, \ce{Ti_{3}C_{2}F_{2}}, \ce{Ti_{3}C_{2}(OH)_{2}}, and \ce{Ti_{2}C_{}F_{2}}.
We study the effect of different surface terminations and monosheet thickness on the electrical conductivity, and show that the changes in conductivity can be explained by the squared velocity density of the electronic state, as well as their phonon scattering lifetime.
We also compare the solution of the iterative Boltzmann transport equation (IBTE) to different linearized solutions, namely, the self-energy relaxation time approximation (SERTA) and the momentum relaxation time approximation (MRTA), and we show that the SERTA significantly underestimates the electrical conductivity while the MRTA yields results in better agreement with the IBTE.
The computed monolayer conductivity at 300K
  is in reasonable agreement with reported
  experimental measurements.

\end{abstract}

\maketitle % Insert title

MXenes form a large family of two-dimensional
  transition metal carbides and nitrides with interesting electrochemical properties
  \cite{leiRecentAdvancesMXene2015a,anasori2DMetalCarbides2017a,naguibTwoDimensionalNanocrystalsProduced2011,naguibTwoDimensionalTransitionMetal2012,enyashinStructuralElectronicProperties2013}.
These layered materials have shown potential for a wide range of applications
  in energy storage and conversion
  \cite{khazaeiRecentAdvancesMXenes2019a,sunTwodimensionalMXenesEnergy2018,yorulmazVibrationalMechanicalProperties2016,xieRoleSurfaceStructure2014,erTiMXeneHigh2014,sunTwodimensionalMXenesEnergy2017a}.
Their high specific surface area and electrochemical activity
  make them suitable for
  supercapacitors \cite{shanTwodimensionalVanadiumCarbide2018,sunTwodimensionalMXenesEnergy2017a,sunTwodimensionalMXenesEnergy2018,rakhiEffectPostetchAnnealing2015,ghidiuConductiveTwodimensionalTitanium2014},
  lithium-ion batteries \cite{naguib25thAnniversaryArticle2014a,naguibMXenePromisingTransition2012,naguibNewTwoDimensionalNiobium2013},
  catalysis \cite{dingPromotingN2Electroreduction2021b},
  photocatalysis \cite{xiePositioningMXenesPhotocatalysis2020,handokoTheoryguidedMaterialsDesign2019},
  and hydrogen storage \cite{lukatskayaMultidimensionalMaterialsDevice2016,huMXeneNewFamily2013}.
With a suitable hydrophilic surface termination,
  MXenes also exhibit electrocatalytic activity for the
  oxygen evolution reaction (OER)
  \cite{handokoTheoryguidedMaterialsDesign2019,tianDensityFunctionalTheory2021a},
  the oxygen reduction reaction (ORR) \cite{yangTailoringElectronicStructure2020a},
  and the hydrogen evolution reaction (HER)\cite{zouSimpleApproachSynthesis2021,chengTwoDimensionalOrderedDouble2018a}.

The terminated MXenes have a general chemical formula
  \ce{M_{n+1}X_{n} T_{x}}
  (n = 1, 2 or 3),
  where M is a transition metal (Sc, Ti, Zr, Hf, V, Nb, Ta, Cr, Mo, etc.),
  X denotes carbon and/or Nitrogen,
  T represents the surface terminations group
  typically -O, -OH or -F \cite{tangMXene2DLayered2018,sunTwodimensionalMXenesEnergy2018,huEmerging2DMXenes2020},
  and x is the number of terminations.
The surface termination of the 2D layers
  originates from their synthesis by chemical etching %
  \cite{wangInvestigationChlorideIon2017,mashtalirIntercalationDelaminationLayered2013,naguib25thAnniversaryArticle2014a,wangEnhancementElectricalProperties2015},
  starting from three-dimensional precursors known as MAX phases \cite{eklund1AXPhasesMaterials2010,tangMXene2DLayered2018},
  of which nearly one hundred compounds have been identified
  \cite{leiRecentAdvancesMXene2015a,anasori2DMetalCarbides2017a,khazaeiElectronicPropertiesApplications2019,
        khazaeiElectronicPropertiesApplications2017a,halimTransparentConductiveTwoDimensional2014}.
Previous first-principles calculations have investigated
  how the surface termination alters the electronic properties of the MXenes
  \cite{sevikSuperconductivityFunctionalizedNiobiumcarbide2022,naguibTwoDimensionalNanocrystalsProduced2011,naguib25thAnniversaryArticle2014a,schultzSurfaceTerminationDependent2019}.
Some MXenes become semiconductors when terminated by oxygen,
    such as \ce{Ti_{2}CO_{2}}, \ce{Zr_{2}CO_{2}}, and \ce{Hf_{2}CO_{2}} \cite{gandiThermoelectricPerformanceMXenes2016},
    while others, like \ce{V_{2}C}, remain metallic for all surface terminations~%
    \cite{champagneElectronicVibrationalProperties2018}.

Beyond their general classification as metals or semiconductors,
  a key property of these materials for most applications is their electrical conductivity.
The electronic transport properties can be computed from first principles
  within the framework of the Boltzmann transport equation (BTE),
  assuming that phonon scattering is the dominant scattering mechanism
  at room temperature and above, and neglecting other scattering channels
  such as defects and impurities
  \cite{dassarmaElectronicTransportTwodimensional2011}.
Furthermore, one avoids solving the BTE iteratively (IBTE)
  by using the self-energy relaxation time approximation (SERTA)
  or the momentum relaxation time approximation (MRTA)
  \cite{liuFirstprinciplesModebymodeAnalysis2017,ponceHoleMobilityStrained2019,poncePredictiveManybodyCalculations2018}.
This framework has been widely used to study the electrical transport semiconductors
  and metals
  \cite{liElectricalTransportLimited2015,ponceHoleMobilityStrained2019,poncePredictiveManybodyCalculations2018,ponceFirstprinciplesCalculationsCharge2020,bruninPhononlimitedElectronMobility2020}.
It has been recently shown, however, 
  that some of these approximations
  may underestimate significantly the charge mobility in semiconductors,
  while the IBTE can be achieved
  at virtually the same computational cost~\cite{claesAssessingQualityRelaxationtime2022}.

In the present work,
  we study the phonon-limited electrical conductivity of four known MXenes:
  \ce{Ti_{3}C_{2}}, \ce{Ti_{3}C_{2}F_{2}}, \ce{Ti_{3}C_{2}(OH)_{2}}, and \ce{Ti_{2}C_{}F_{2}}.
The comparison among these materials
  allows us to discern the influence of 
  surface termination and monosheet thickness
  on the scattering lifetime of the charge carriers.
We also compare the different frameworks
  for computing the electrical conductivity,
  and find that the conclusions of Claes~et.~al~\cite{claesAssessingQualityRelaxationtime2022}
  do hold for this class of two-dimensional metallic systems,
  namely that 
  the SERTA approach underestimates the electrical conductivity
  while the MRTA results are in better agreement with the IBTE.
We show that the predicted conductivity is consistent
    with experimental measurements.

\begin{figure*}[t]
\begin{tabular}{c c c c}
    \ce{Ti_{3}C_{2}}
  & \ce{Ti_{3}C_{2}(OH)_{2}}
  & \ce{Ti_{3}C_{2}F_{2}}
  & \ce{Ti_{2}CF_{2}}
  \\
	\includegraphics[width=0.18\textwidth]{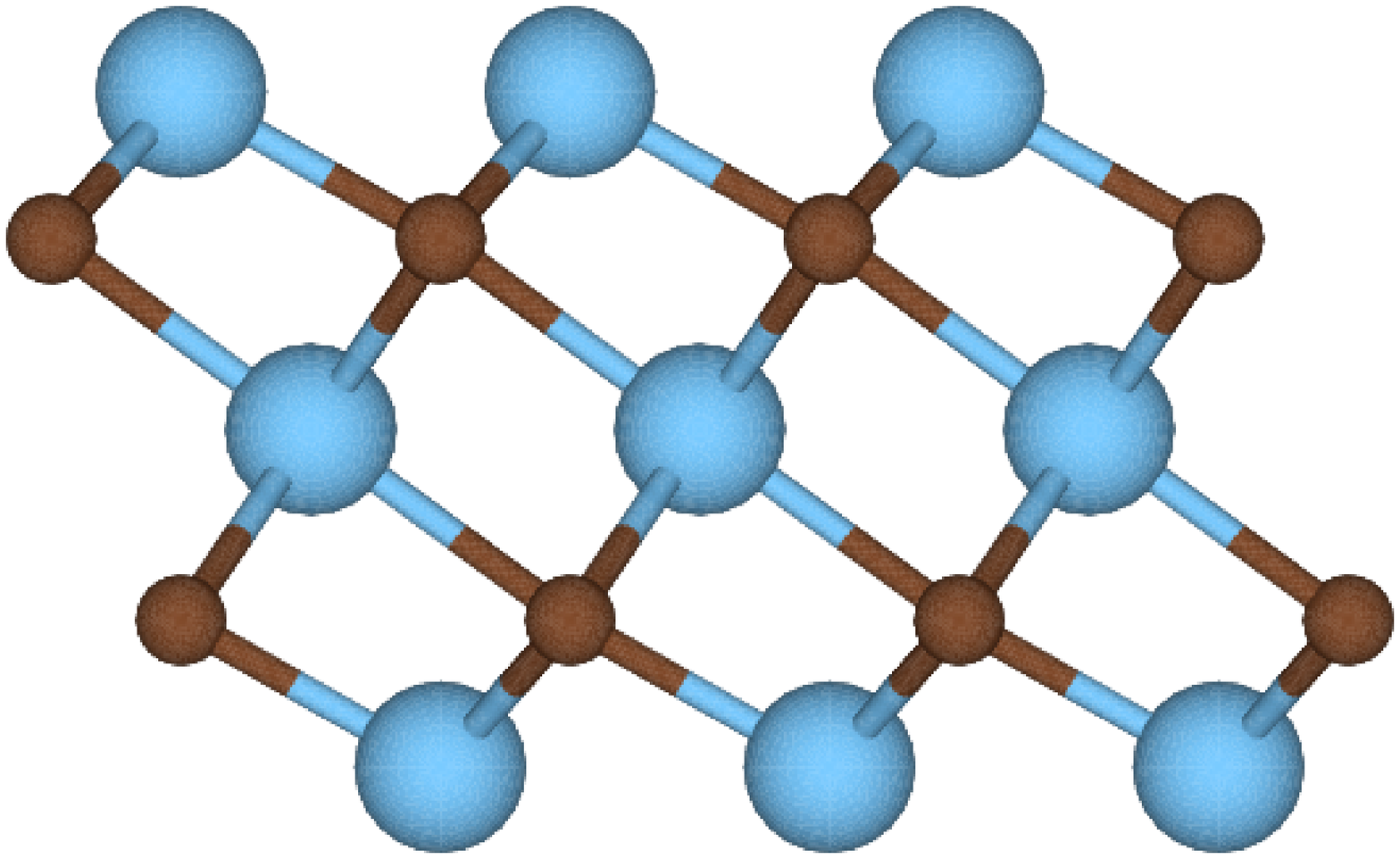}
  &
	\includegraphics[width=0.18\textwidth]{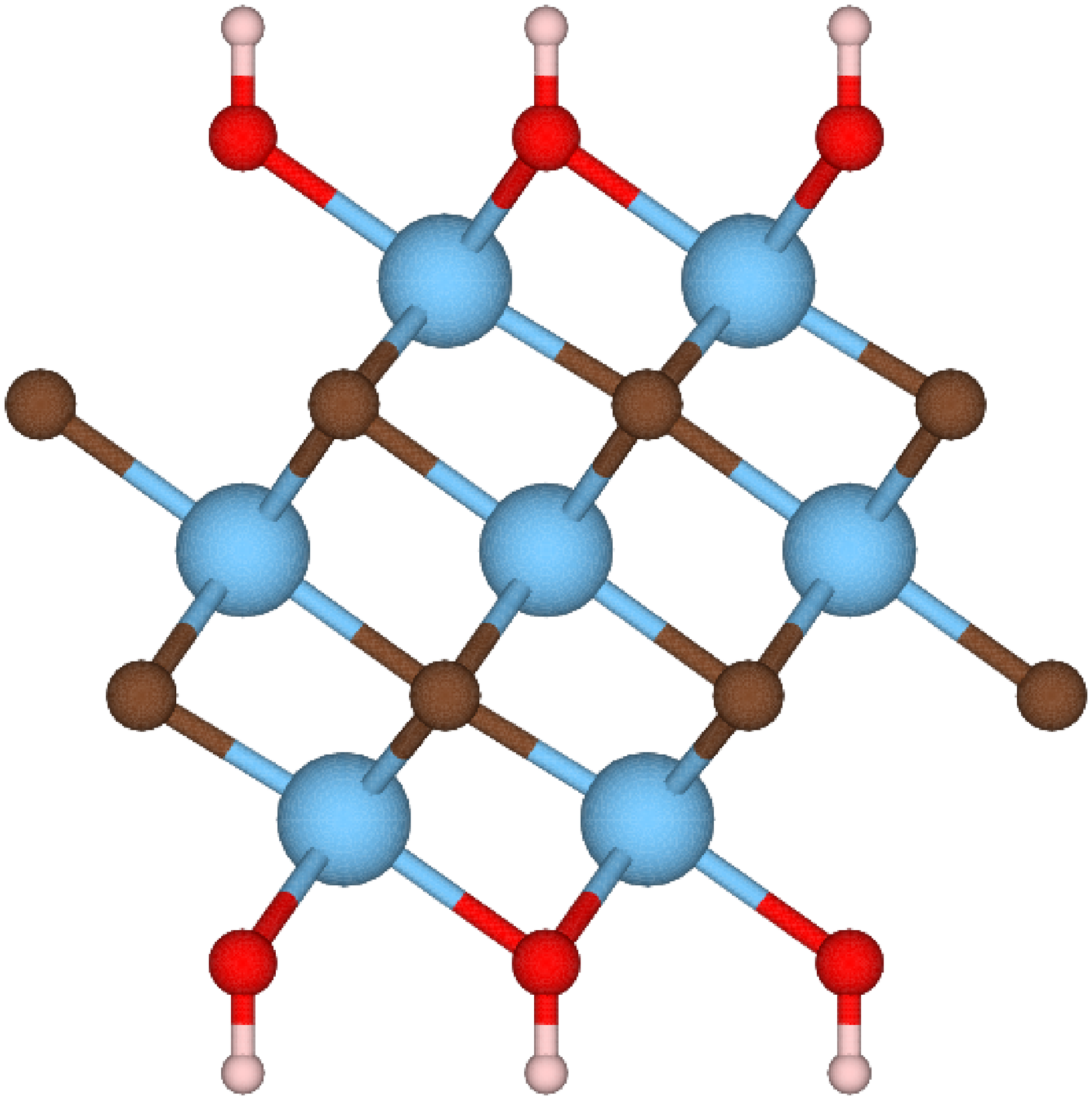}
  &
	\includegraphics[width=0.18\textwidth]{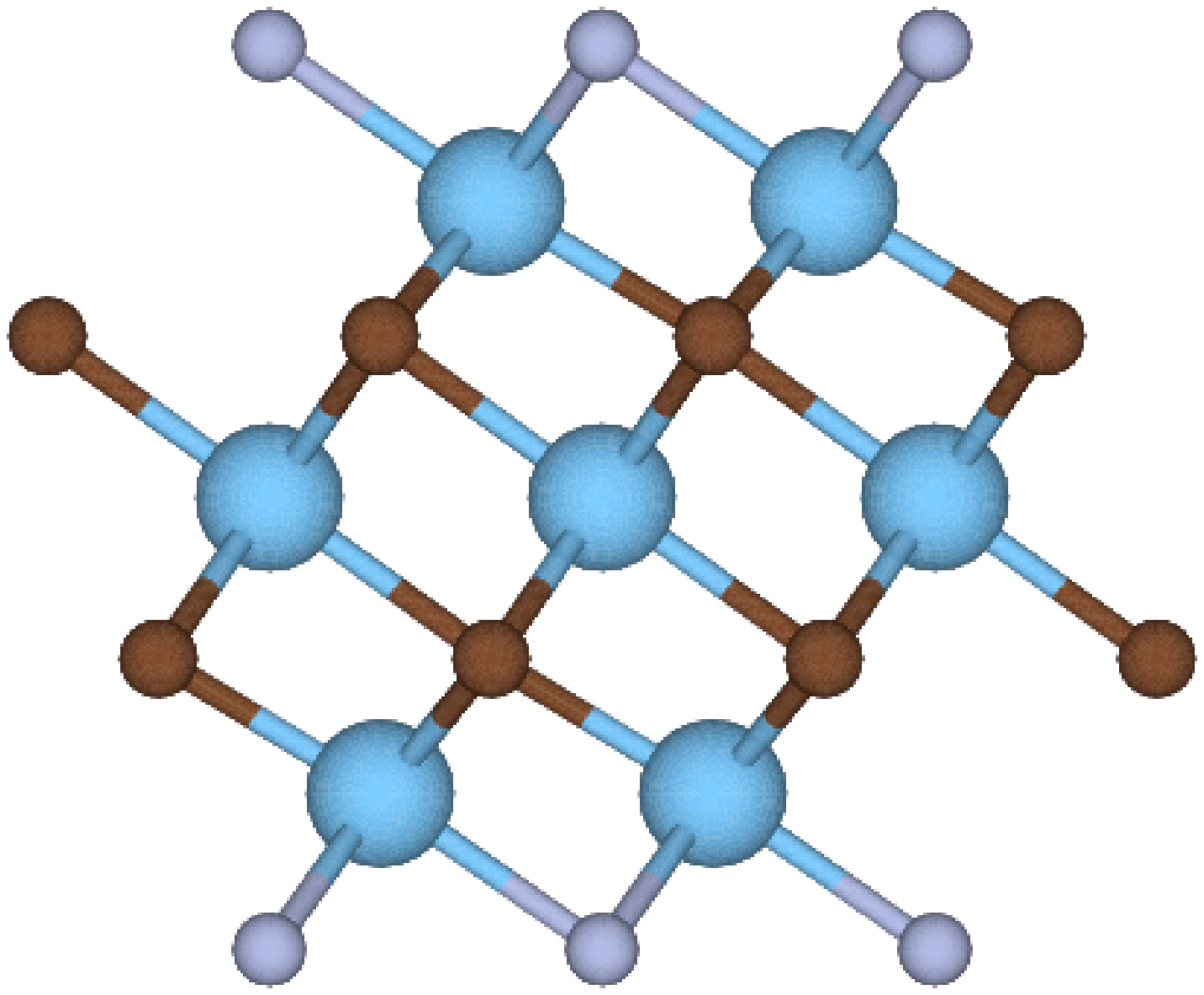}
  &
	\includegraphics[width=0.18\textwidth]{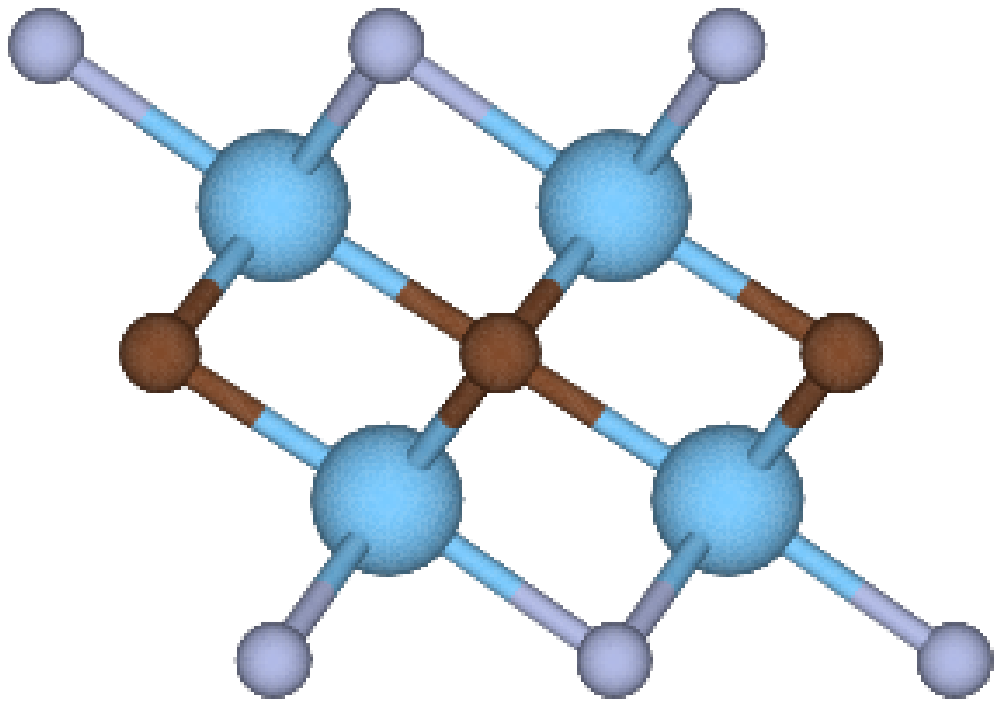}
  \\
  \bf a & \bf b & \bf c & \bf d
  \\
	\includegraphics[width=0.24\textwidth]{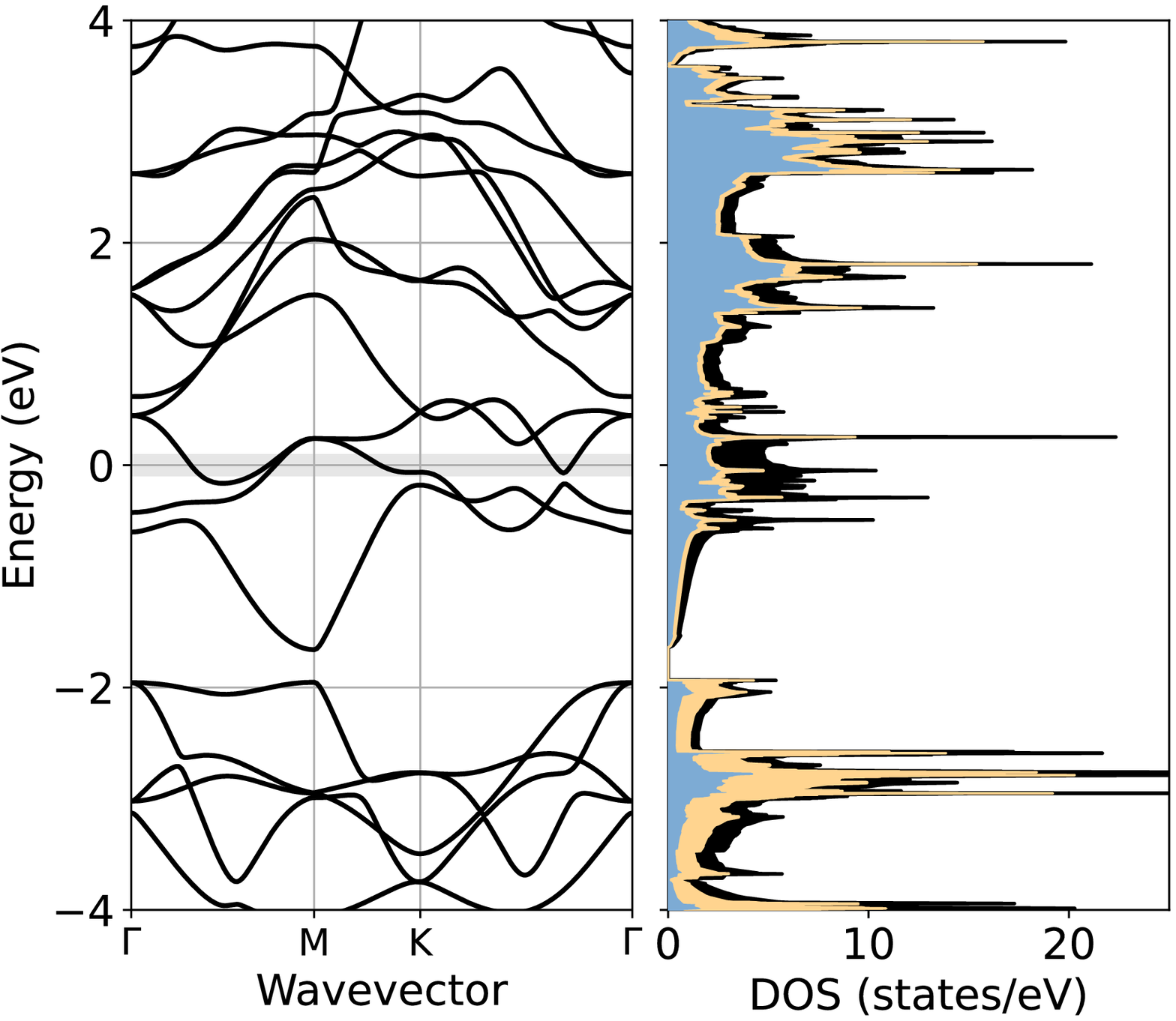}
  &
	\includegraphics[width=0.24\textwidth]{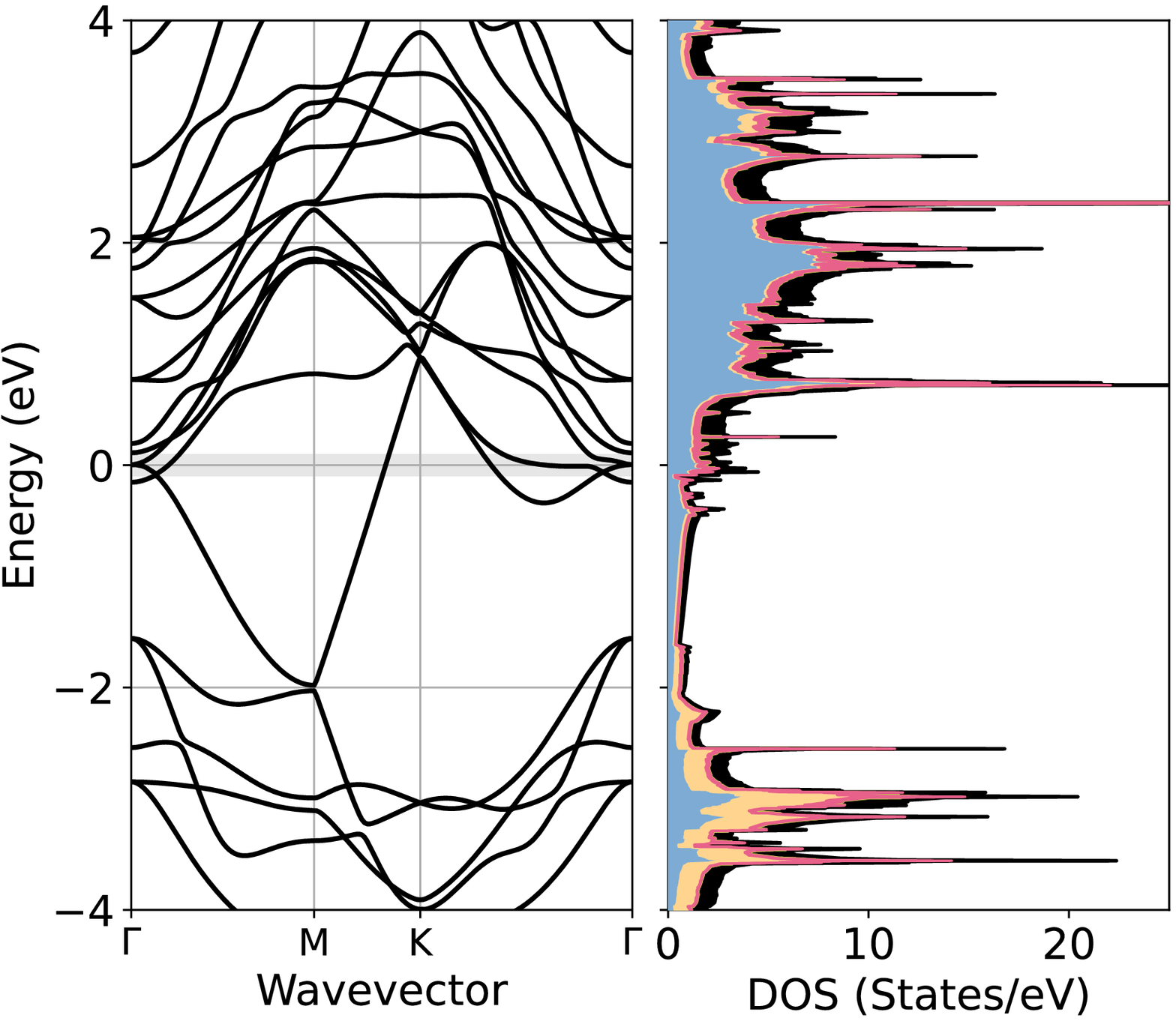}
  &
	\includegraphics[width=0.24\textwidth]{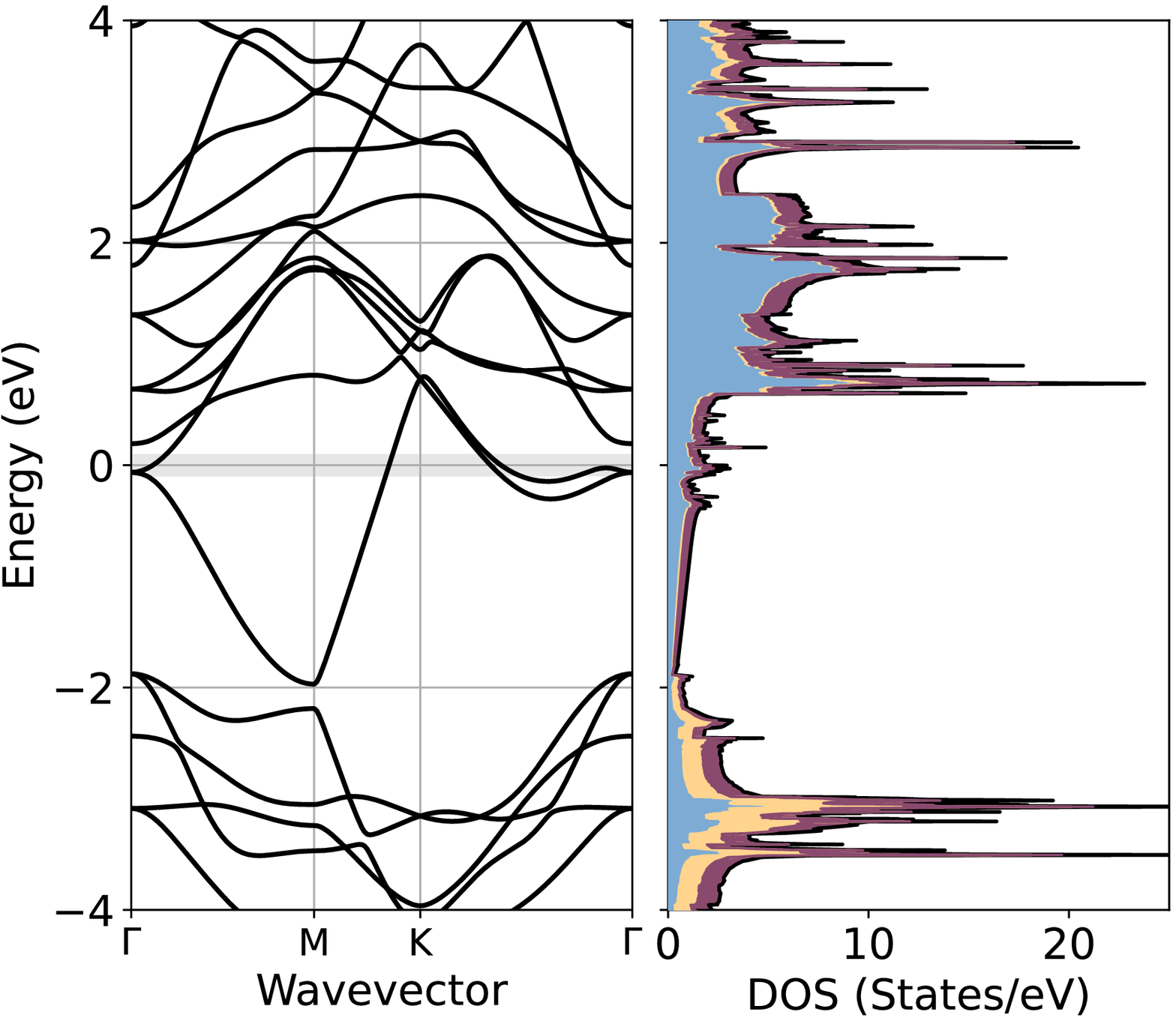}
  &
	\includegraphics[width=0.24\textwidth]{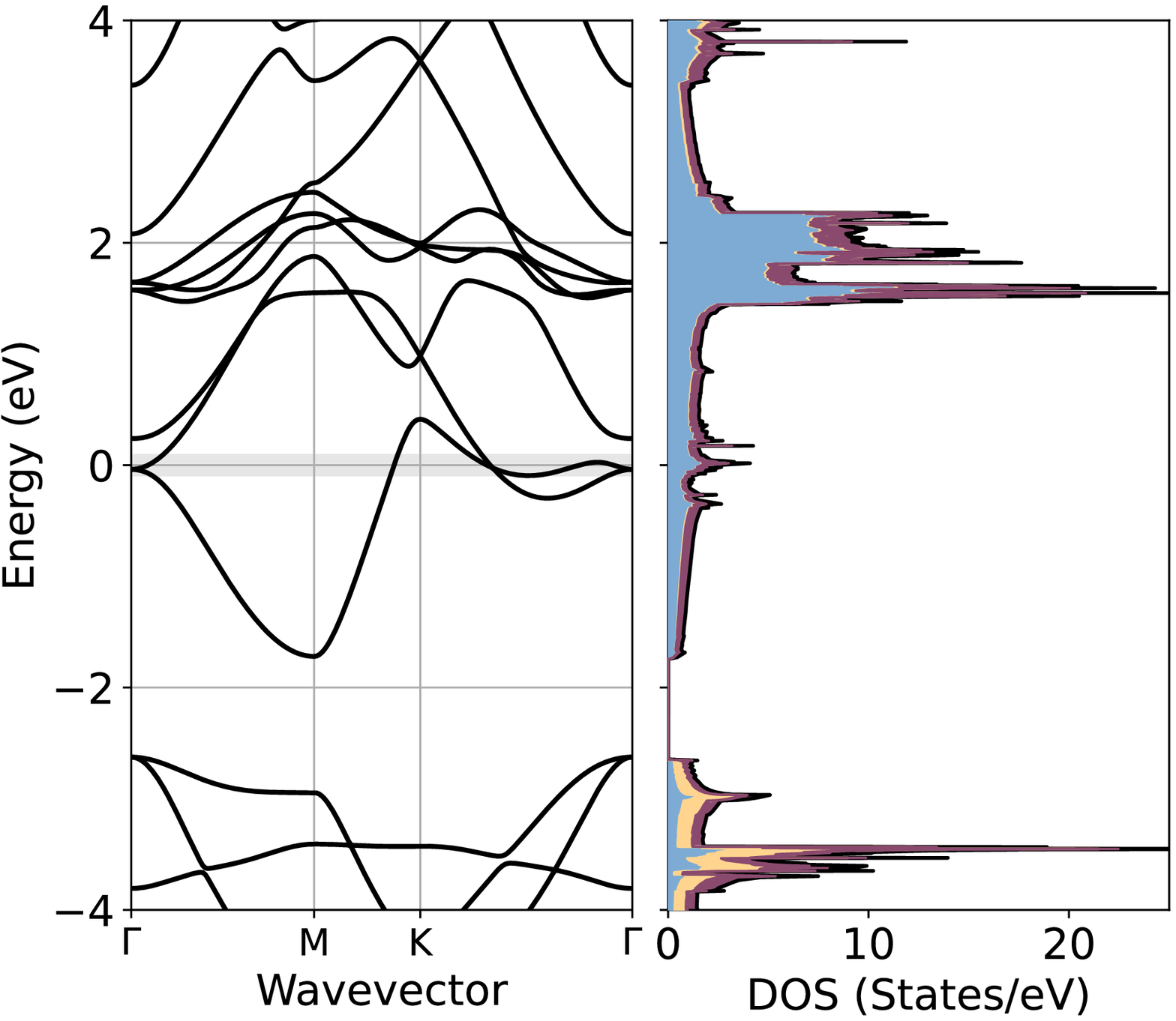}
  \\
  \bf e & \bf f & \bf g & \bf h
  \\
	\includegraphics[width=0.24\textwidth]{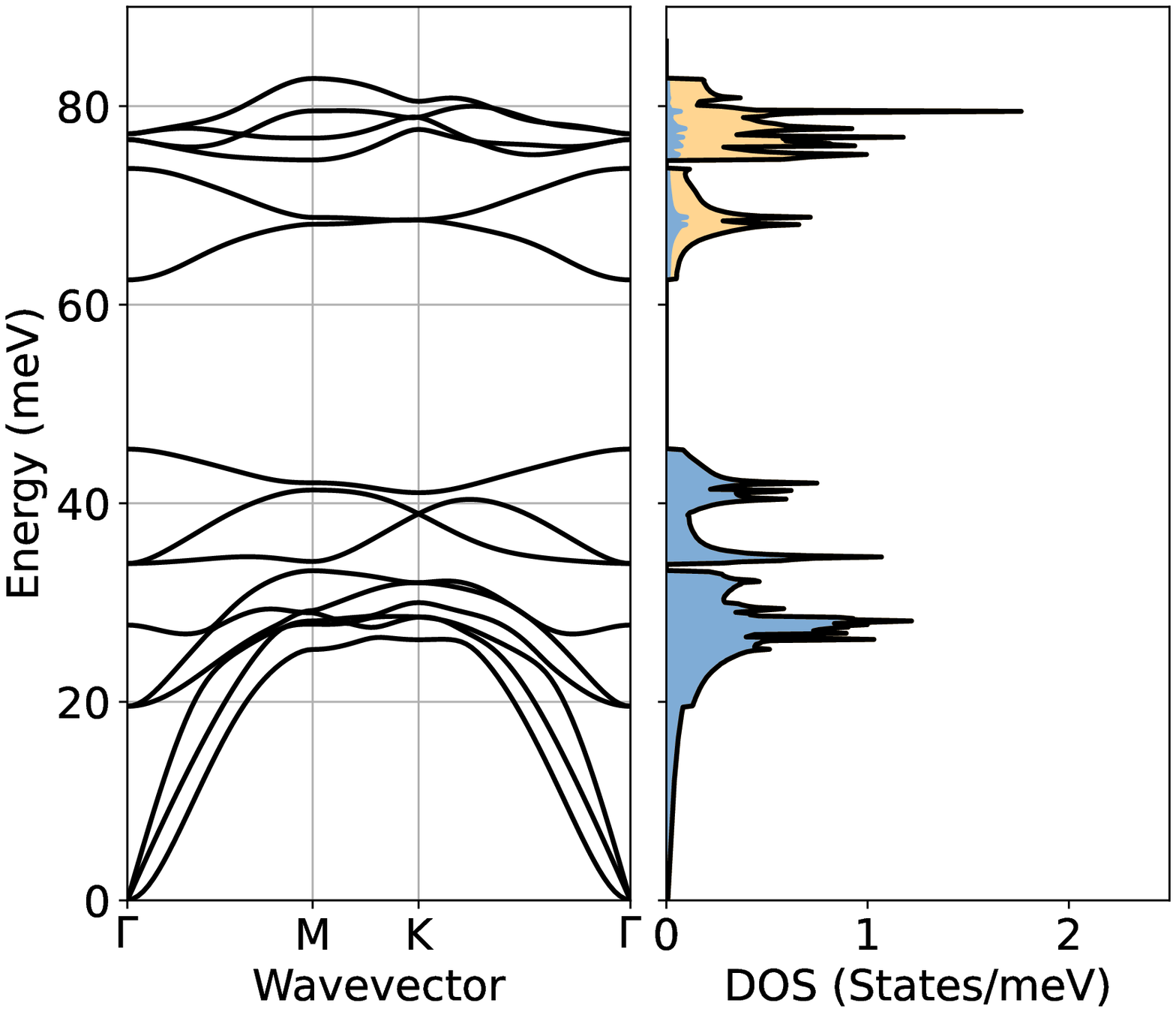}
  &
	\includegraphics[width=0.24\textwidth]{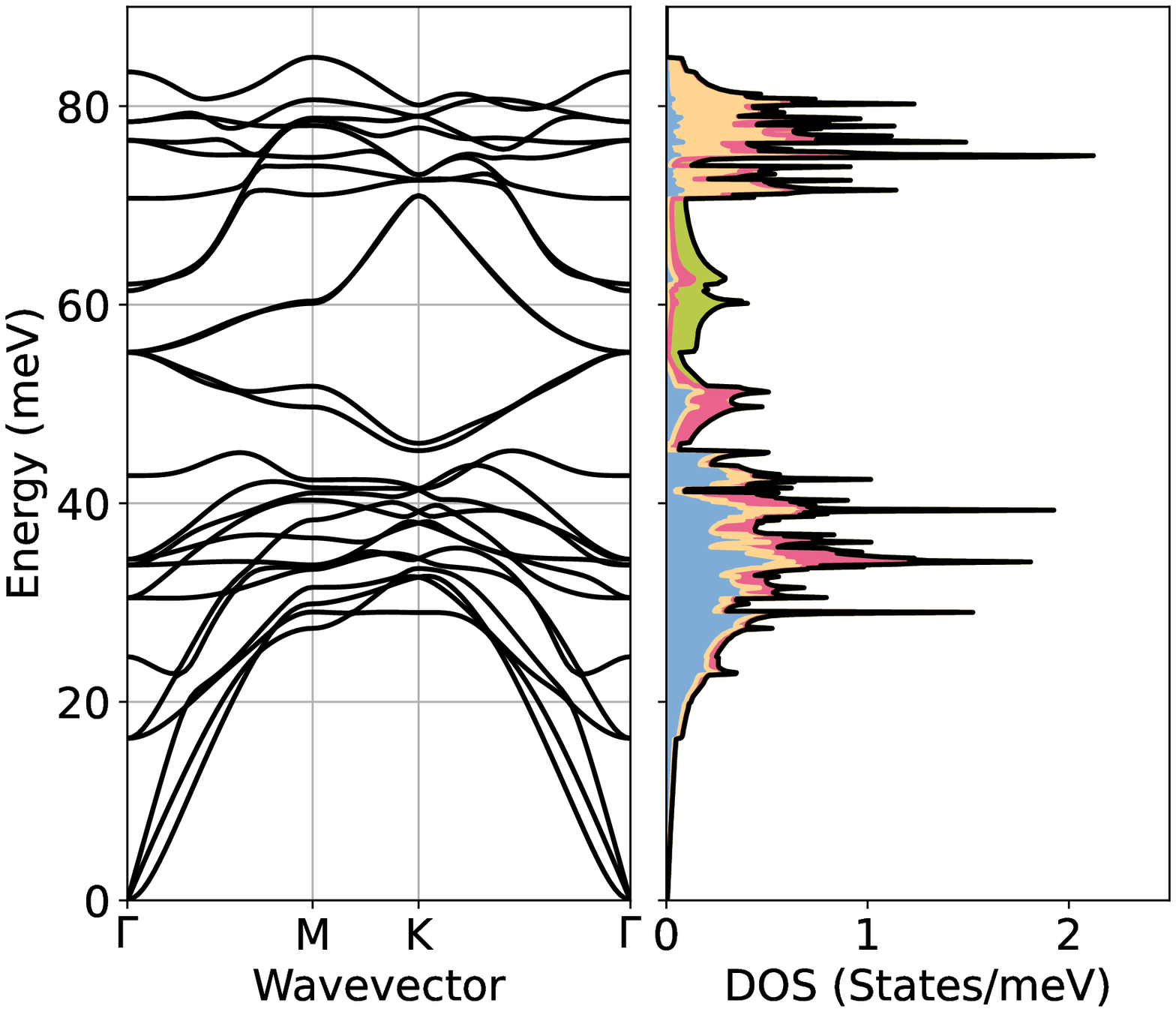}
  &
	\includegraphics[width=0.24\textwidth]{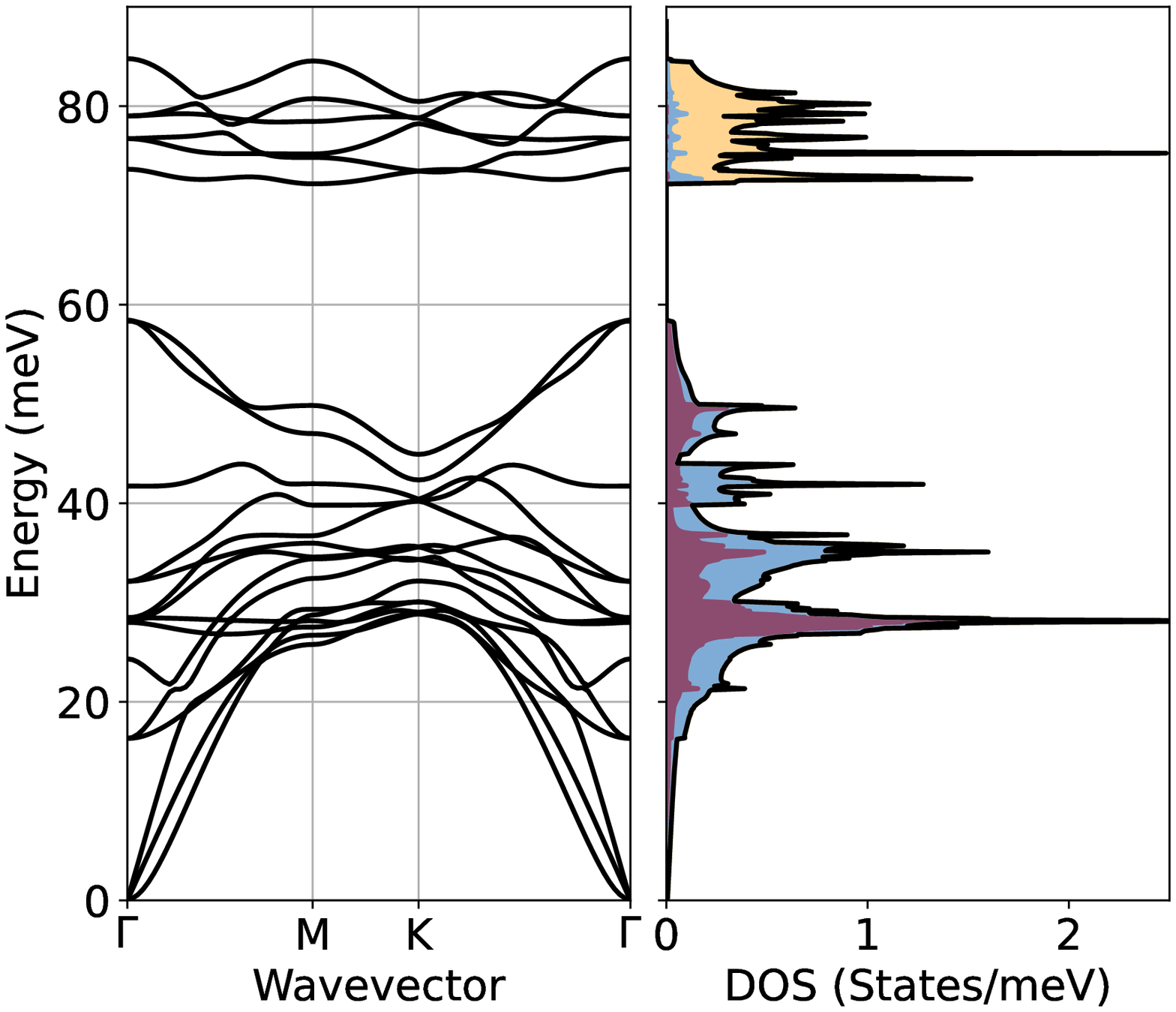}
  &
	\includegraphics[width=0.24\textwidth]{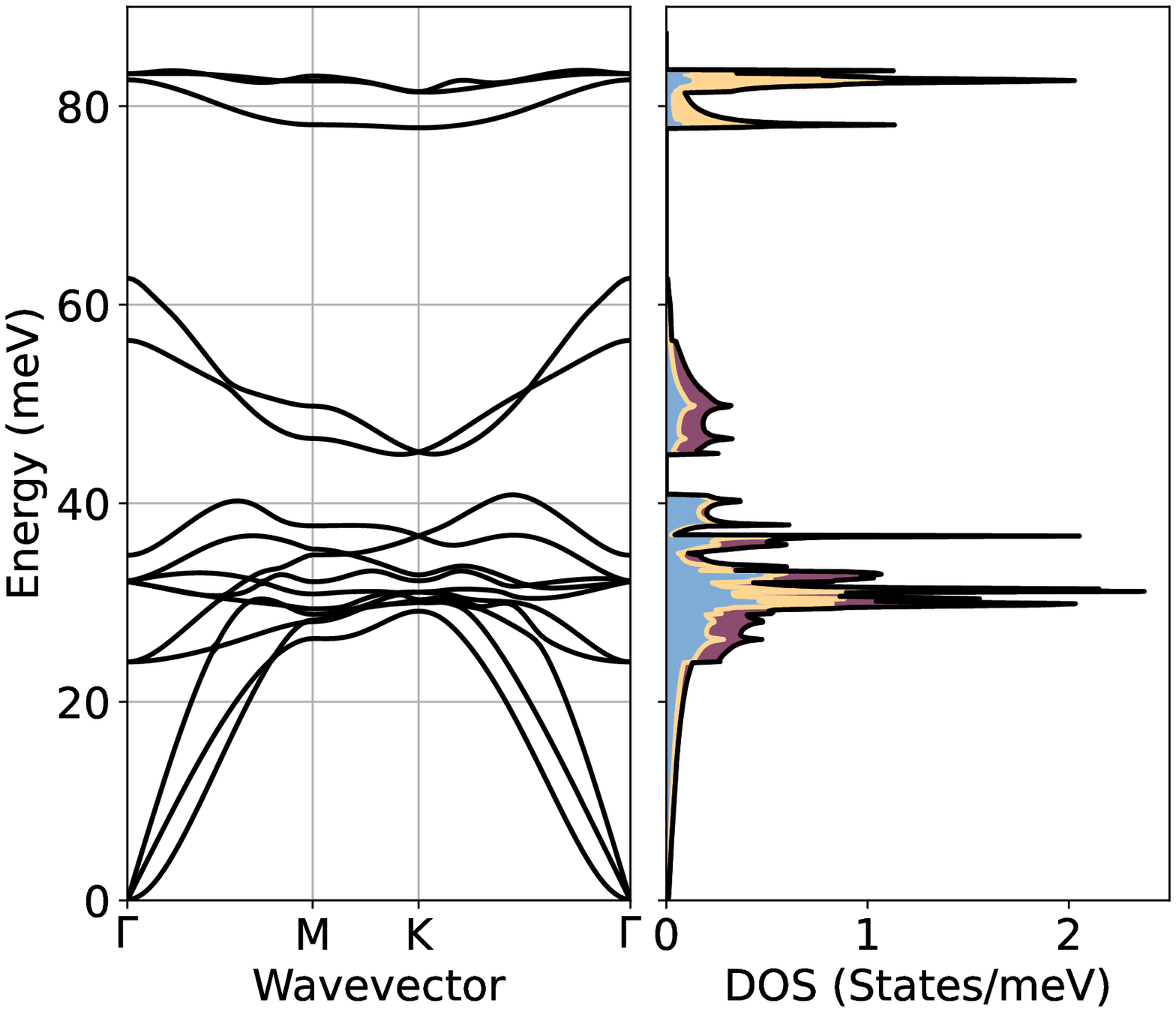}
  \\
  \bf i & \bf j & \bf k & \bf l
  \\
\end{tabular} \\
\includegraphics[width=\textwidth]{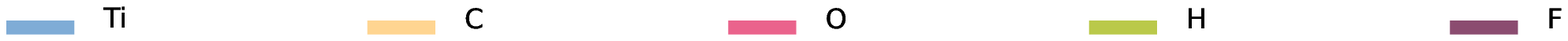}
\vspace{-10mm}
	\caption{ \label{fig:structures_bands_phonons}
  \textbf{a-d}
  Structure of the monosheet materials.
  Titanium atoms are in blue, carbon in brown, fluorine in grey, oxygen in red and hydrogen in pink.
  \textbf{e-h}
	  Electronic band structures along the high symmetry directions and Projected Density of States (PDOS).
    The energy levels are referenced to the Fermi level at zero.
  \textbf{i-l}
	Phonon band structures along the high symmetry directions and Projected Phonon Density of States.
    Note that \ce{Ti_{3}C_{2}(OH)_{2}} also possess a two phonon branches
    associated with the motion of the hydrogen atoms
    at an energy of 449~meV.
  }
\end{figure*}

\section{Results and discussion}

\subsection{Computational details}

We perform density functional theory (DFT)
  and density functional perturbation theory (DFPT) calculations
  using the Abinit software \cite{gonzeABINITFirstprinciplesApproach2009,gonzeAbinitprojectImpactEnvironment2020},
  to obtain the structural, electronic, and vibrational properties
  of the materials.
We use the PBE exchange correlation functional \cite{perdewGeneralizedGradientApproximation1996},
  with pseudopotentials from the Pseudo-Dojo database~\cite{vansettenPseudoDojoTrainingGrading2018}.
For all the structures considered,
  we use an energy cutoff of 35 Hartree to represent the wavefunctions,
  a $16\times16\times1$ grid of k-points to sample the Brillouin zone,
  such that the energy is converged within $10^{-6}$ eV/cell.
We perform geometry optimization to relax the forces below $10^{-5}$ eV/$\AA$.
A vacuum spacing of 20~\AA is used, in order to avoid interactions between the periodic images of the MXenes monolayers.

\subsection{Structural parameters}

All the 2D materials considered assume the space group $P6_{3}/mmc$.
We perform geometry optimization to relax the forces below $10^{-5}$ eV/$\AA$
  and obtain the relaxed lattice parameters of
  3.098~\AA, 3.076~\AA, 3.086~\AA, and 3.059~\AA~for the
  \ce{Ti_{3}C_{2}}, \ce{Ti_{3}C_{2}F_{2}}, \ce{Ti_{3}C_{2}OH_{2}}, and \ce{Ti_{2}CF_{2}}
  monosheets, respectively.
Several possible configurations exist for the surface termination of \ce{Ti_{n+1}C_{n}T_{x}}~%
  \cite{tangAreMXenesPromising2012a,baiDependenceElasticOptical2016,huEmerging2DMXenes2020}.
We use the most energetically favorable configuration
  where surface termination atoms (\ce{F} or \ce{OH}) are at the hollow site
  of three neighboring carbon atoms.
The atomic structure relaxations of all the materials considered
  are presented in Fig~\ref{fig:structures_bands_phonons}(a-f).

\subsection{Electronic bands}

The band structure and the projected density of states (PDOS)
  of the materials are presented in Fig~\ref{fig:structures_bands_phonons}(e-h).
All the materials are metallic,
  with the electronic states near the Fermi level
  mostly composed of Ti $d$ orbitals.
For the surface-terminated systems,
  the band in the $M-K$ direction is highly dispersive at the Fermi level,
  suggesting a high electrical conductivity \cite{wangFirstprinciplesStudyStructural2014}.
We note the presence of a valley and a flat band region
  along the $\Gamma-K$ direction,
  which contribute to singularities in the density of states
  and represent a potential scattering channels for the charge carriers.

\subsection{Phonon bands}

In Fig~\ref{fig:structures_bands_phonons}(i-l),
  we present the phonon band structures and the projected phonon density of states. % (PJDOS).
These results were obtained by employing a coarse q-points meshes of $8\times8\times1$ for $Ti_{3}C_{2}$, $Ti_{3}C_{2}F_{2}$ and $T_{3}C_{2}(OH)_{2}$, and $16\times16\times1$ for $Ti_{2}CF_{2}$.
Every phonon frequency is real and positive,
indicating that the structures are stable
with respect to atomic displacements \cite{khazaeiOHterminatedTwodimensionalTransition2015,gholivandEffectSurfaceTermination2019}.

From the projected phonon density of states,
    we see a clear energy separation between the
    phonon modes associated with the different atomic species.
The low frequency bands correspond to 
    the vibrating motion of the metallic atoms,
    the high frequency bands
    are associated with the motion of carbon atoms,
    and the surface terminations bring additional phonon bands
    at intermediate energies.
This general feature has been observed in other MXene materials as well~\cite{yorulmazVibrationalMechanicalProperties2016,champagneElectronicVibrationalProperties2018}.

\subsection{Electrical conductivity}

The electrical conductivity can be computed by solving the iterative Boltzmann transport equation (IBTE)~\cite{poncePredictiveManybodyCalculations2018,brown-altvaterBandGapRenormalization2020,giustinoElectronphononInteractionsFirst2017}
where the main scattering mechanism
are the phonon collisions,
which are described by the electron-phonon coupling matrix elements computed from first principles.
By making use of the relaxation time approximation,
    the BTE can be linearized, avoiding the iterative procedure
    and writing the electrical conductivity $\sigma_{\alpha}$ as
\begin{equation} \label{eq:conductivity}
    \sigma_{\alpha} \  = \frac{-e}{\Omega}
    \sum_n \int \frac{d\mathbf{k}}{\Omega_{BZ}}
    \tau_{n\mathbf{k}}(T)
    \lvert v_{n\mathbf{k}\alpha} \rvert^2
    f'(\varepsilon_{n\mathbf{k}})
\end{equation}
where
  $\alpha$ is a Cartesian direction,
  $\Omega$ is the volume of the unit cell,
  $\Omega_{BZ}$ is the volume of the Brillouin zone,
  $\tau_{n\mathbf{k}}(T)$ is the temperature-dependent scattering lifetimes
  of the electron state,
  $v_{n\mathbf{k}\alpha}$ is the electron velocity,
  and $f'(\varepsilon)$ is the derivative of the Fermi-Dirac distribution, which depends on temperature.
Different approximations exist for the computation of the electron scattering lifetime, including the Self-Energy Relaxation Time Approximation (SERTA) \cite{ponceHoleMobilityStrained2019,poncePredictiveManybodyCalculations2018}
    and the Momentum Relaxation Time Approximation (MRTA) \cite{liElectricalTransportLimited2015,claesAssessingQualityRelaxationtime2022}.
In the former, the lifetime is computed from the inverse of the Fan-Migdal
    electron-phonon coupling self-energy,
    and in the latter, the lifetime is computed from the squared electron-phonon coupling
    matrix elements weighted by an efficiency factor
    which accounts for the momentum direction of the scattering states
    relative to the electrical field~\cite{claesAssessingQualityRelaxationtime2022}.

\begin{figure}
	\includegraphics[width=\linewidth]{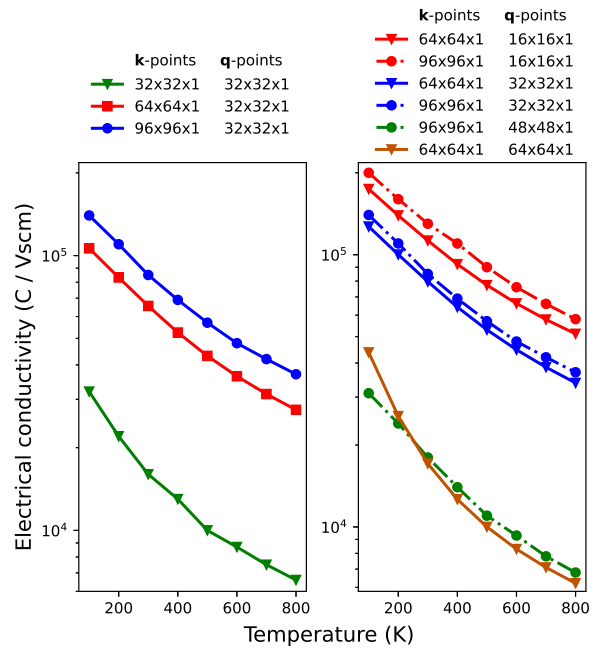} 
	\caption{
    Convergence of the electrical conductivity in \ce{Ti_{3}C_{2}F_{2}}
    with respect to the Brillouin zone sampling
    of electron (\textbf{k}) and phonon (\textbf{q}) wavevectors.
    Left:  Varying \textbf{k}-points grids and fixed \textbf{q}-points grid.
    Right: Varying \textbf{q}-points grids and commensurate \textbf{k}-points grids.
	} \label{Convergence Study}
\end{figure}

\subsection{Convergence study}

One of the main challenges in computing the electrical conductivity is the fine sampling of
electron (\textbf{k}-points) and phonon (\textbf{q}-points) wavevectors
required to converge the transport properties \cite{gonzeAbinitprojectImpactEnvironment2020,wenLargeLatticeThermal2020}.
A dense \textbf{k}-mesh is required to achieve good sampling of the electronic occupations near the Fermi level, while a dense \textbf{q}-sampling is required to converge the electronic lifetimes \cite{wenLargeLatticeThermal2020}.
This is especially true in two-dimensional metals, where the density of states is expected to vary rapidly near the Fermi level, as can be seen in Fig \ref{fig:structures_bands_phonons}(\textbf{e-h}).

In order to optimize the overall computational cost,
    we employ the Shankland-Koelling-Wood interpolation scheme \cite{shanklandFourierTransformationSmooth2009,koellingInterpolationEigenvaluesResultant1986},
    a feature recently made available
    within the Abinit automated workflows~\cite{gonzeAbinitprojectImpactEnvironment2020,bruninPhononlimitedElectronMobility2020}.
The electronic energies and wavefunctions
    near the Fermi level are interpolated from coarse \textbf{k}-grid
    onto a fine \textbf{k}-grid.
We set the coarse \textbf{k}-grid to $16\times16\times1$
    and vary the fine \textbf{k}-grid to perform the convergence study.

Figure~\ref{Convergence Study} shows the convergence of the temperature-dependent electrical conductivity of \ce{Ti_{3}C_{2}F_{2}} with varying \textbf{k}-points and \textbf{q}-points grids.
From this figure, we conclude that a $64\times64\times1$ homogeneous grid for both \textbf{k}-points and \textbf{q}-points are sufficiently converged, and we use these parameters for all the studied materials.

\subsection{Results for the electrical conductivity}

\begin{figure}
	\includegraphics[width=\linewidth]{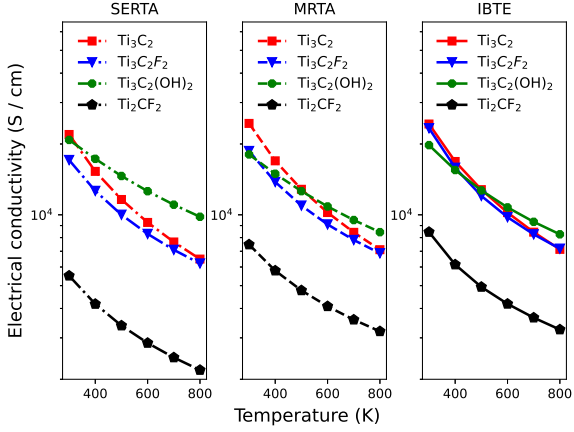} 
	\caption{
    Temperature-dependent electrical conductivity of each structures
    with the linearized (SERTA, MRTA) and iterative Boltzmann transport equation (IBTE).
    }\label{Electrical conductivity}
\end{figure}

The temperature-dependent electrical conductivity of the four MXenes with the different approximations is presented in Fig.~\ref{Electrical conductivity}.
We note, again, that the IBTE calculation uses the same computational cost as
    the  SERTA and MRTA calculations.
We find %, in agreement with the trend identified by Claes et al.~\cite{claes2022assessing},
    that the SERTA  underestimates the conductivity by as much as 14 \% at T=300 K and 4 \% at T=800 K, compared to the IBTE, while the MRTA is in somewhat better agreement with the IBTE (9 \% at T=300 K and 0.8 \% at T=800 K).

\begin{figure}
	\includegraphics[scale=0.55]{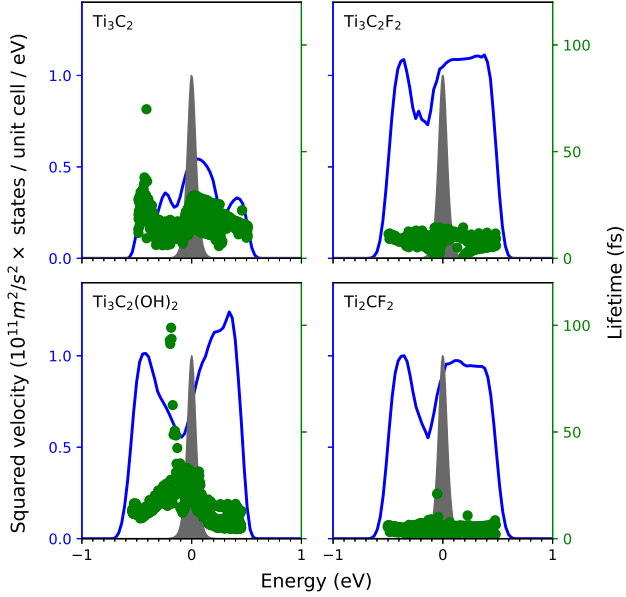} 
	\caption{
    The phonon self-energy lifetime at $T=300K$ of each electronic state
    near the Fermi level (green discs)
    and the squared velocity density (blue)
    with $\alpha=1$, that is, the direction along a primitive vector.
    The grey shaded curve represents the negative of the derivative of the Fermi-Dirac distribution.
	}
    \label{fig:v2tau}
\end{figure}

In Figure~\ref{fig:v2tau},
    we decompose the SERTA electrical conductivity
    into functions separating the integrants of  Eq.~\eqref{eq:conductivity}.
The derivative of the equilibrium Fermi Dirac distribution function
$- \frac{\partial f}{\partial \varepsilon}$
    is peaked around the Fermi level
    and indicates the energy range where the electronic states
    may contribute to the conductivity.
The squared velocity density $\langle v^2 \rangle$ is defined as
\begin{equation}
    \langle v_{\alpha}^2\ (\varepsilon) \rangle = \sum_n \int \frac{d\mathbf{k}}{\Omega_{Bz}} \lvert {v_{n\mathbf{k}\alpha} \rvert}^2\
    \delta(\varepsilon-\varepsilon_{n\mathbf{k}})
\end{equation}
This function is indicative of the number of carriers available
    for conductivity.
It is a smooth function of energy, unlike the density of states,
    which, for 2D materials, has spiky structure
    that requires a high number of k-points to sample.
By comparing the $\langle v^2 \rangle$ function of the different materials,
    we note that the addition of either surface termination to the
    $Ti_3C_2$ sheet results in an increase
    of the squared velocity density.
This is due to the surface termination atoms
    pulling electrons off the central layer
    and lowering the Fermi level
    to intercept the highly dispersive Ti $d$-band
    along the $M-K$ segment of the bandstructures.

Looking at the electronic scattering lifetimes $\tau_{n\mathbf{k}}$
    shown in Fig~\ref{fig:v2tau},
    we note that \ce{Ti_{2}CF_{2}},
    being the thinnest monosheet,
    also has the shortest lifetimes.
Among the terminated structures,
    \ce{Ti_{3}C_{2}F_{2}} and \ce{Ti_{2}CF_{2}}
    have shorter scattering lifetimes
    than \ce{Ti_{3}C_{2}(OH)_{2}},
    most likely due to the presence of
    flat bands near the Fermi level in the $\Gamma-K$ region
    for the fluorinated structures.
As a result, according to the SERTA calculation,
     \ce{Ti_{3}C_{2}(OH)_{2}} has
     the highest electrical conductivity.
However, the full IBTE calculation reveals instead that
    \ce{Ti_{3}C_{2}F_{2}} has a higher electrical conductivity
    than \ce{Ti_{3}C_{2}(OH)_{2}}
    at temperatures up to 500~K.

\subsection{Comparison with experiments}

Several conductivity measurements of layered \ce{Ti_{3}C_{2}T_x}
  are reported
  with a variety of experimental setups 
  \cite{chenPristineTitaniumCarbide2020,halimTransparentConductiveTwoDimensional2014b,hoSensingMXenesProgress2021,iqbalMXenesElectromagneticInterference2021, lingFlexibleConductiveMXene2014,lipatovHighElectricalConductivity2021,liuMXenesOptoelectronicDevices2021,m.w.barsoumSynthesisLowCostTitaniumBased2016,mathisModifiedMAXPhase2021, mirandaElectronicPropertiesFreestanding2016, mirkhaniHighDielectricConstant2019, naguibTenYearsProgress2021, qiaoElectricalConductivityEnhancement2021, sarycheva2DTitaniumCarbide2018, shahzad2DTransitionMetal2020, shahzadElectromagneticInterferenceShielding2016, shayestehzeraatiImprovedSynthesisTi2021, songMXenesPolymerMatrix2021, tangInterlayerSpaceEngineering2021, zhangScalableManufacturingFree2020, zhangScalableManufacturingFree2020a}.
The surface termination is either \ce{F} or \ce{OH}, but is generally unspecified.
In practice, the bulk conductivity of MXene flakes depends
  on the synthesis method, which may yield different concentrations of defects and impurities,
  as well as different spacings between nanosheets.

From a dimensional analysis, the bulk conductivity must be proportional
  to the density of MXene nanosheets
  as $\sigma = \sigma_{2D} / L_z$,
  where $\sigma_{2D}$ is the monolayer conductivity 
  with units of Siemens (S),
  and $L_z$ is the interlayer distance.
In the present calculation, $L_z$ is set arbitrarily to 20~$\AA$,
  wereas in experiments, $L_z$ is inferred from XRD spectra.

Table~\ref{tab:expcomp} presents a list
  of recent experimental measurements
  of electrical conductivities. 
While the experimental precision is on the fourth significant digit or better,
    the measured values among different samples
    vary by an order of magnitude.
The present calculation corresponds to a defect-free system,
  in which electron-phonon scattering is the only source of resistivity,
  and sets an upper bound on the conductivity.
Comparing
  against the highest electrical conductivity achieved
  for \ce{Ti_{3}C_{2}T_{x}},
  the computed IBTE value at 300K
  is indeed larger by 59\% for \ce{Ti_{3}C_{2}F_{2}}
  and 34\% for \ce{Ti_{3}C_{2}(OH)_{2}}.

\begin{table}[H]
    \begin{ruledtabular}
	  \setlength{\tabcolsep}{1pt}
    \begin{tabular}{c c c c c}
      Material
      & $\sigma$ ($10^{3}$ S$\cdot$cm$^{-1}$)
      & $L_{z}$ (\AA)
      & $\sigma_{2D}$ ($10^{-3}$ S)
      & Reference
    \\
    \midrule
        \ce{Ti_{3}C_{2}T_x}
      & $24.0$
      & $12.34$
      & $2.96$
      & \cite{shayestehzeraatiImprovedSynthesisTi2021}
    \\
     & $5.8$
     & $11.94$
     & $0.69$
     & \cite{shayestehzeraatiImprovedSynthesisTi2021}
    \\
     & $15.0$ %$ \pm 160$
     & $11.5$
     & $1.73$  % $\pm 0.0184 \times 10^{-3}$
     & \cite{zhangScalableManufacturingFree2020a}
    \\
     & $10.5$ %$ \pm 0.24$
     & $13.8$
     & $1.45$  %$\pm 0.03312$
     & \cite{zhangScalableManufacturingFree2020a}
    \\
     & $11.0$
     & $12.5$
     & $1.38$
     & \cite{lipatovHighElectricalConductivity2021}
    \\
     & $10.4$ %$\pm 200$
     &  $10.4$
     &  $1.08$
     & \cite{chenPristineTitaniumCarbide2020}
    \\
    \midrule
       \ce{Ti_{3}C_{2}F_2}
     & 23.33
     & $20.0$
     & 4.67
     & This work
    \\
       \ce{Ti_{3}C_{2}(OH)_2}
     & 19.78
     & $20.0$
     & 3.96
     & 
    \\
    \end{tabular}
    \end{ruledtabular}
    \caption{\label{tab:expcomp}%
      Measured and computed electrical conductivity at 300K:
      bulk conductivity ($\sigma$),
      interlayer distance ($L_z$),
      monolayer conductivity ($\sigma_{2D}$).      
    }
\end{table}

Aside from the \textbf{k}-points / \textbf{q}-points convergence,
  and unaccounted scattering channels like defects,
  other sources of error in our theoretical calculation
  include the thermal expansion of the lattice,
  the renormalization of the electron velocities due to phonons~\cite{brown-altvaterBandGapRenormalization2020},
  as well as the accuracy of the exchange-correlation functional
  for the band strucutre~\cite{poncePredictiveManybodyCalculations2018}
  and the electron-phonon coupling strength~\cite{Antonius2014,karsai_electron-phonon_2018,li_electron-phonon_2019}.
Overall, an overestimation by about 50\%
  represents a reasonably good agreement,
  and an accuracy comparable
  to that of typical mobility calculations in 2D materials%
  ~\cite{ponceFirstprinciplesCalculationsCharge2020}.

\section{Conclusion}

In summary, we performed a comparative study of the electronic transport
  of the pristine \ce{Ti_{3}C_{2}},
  terminated \ce{Ti_{3}C_{2}T_{2}} (T = F, OH)
  and \ce{Ti_{2}CF_{2}} monosheets from first principles.
We computed the electrical conductivity of the MXenes %on a homogeneous grid 
  with different relaxation time approximations,
  as well as with the iterative Boltzmann transport equation.
We found that the SERTA underestimates the conductivity,
    while the MRTA is in better agreement with the IBTE.
However, the relative differences among monosheets with different surface terminations can only be resolved by the iterative procedure.
Nonetheless,
    the relaxation-time approximation provides a useful understanding
    of the underlying physics
    by decomposing the electrical conductivity
    into scattering lifetime and squared velocity density.
The computed monolayer conductivity
  is in reasonable agreement with experiments,
  within 30\% to 60\%.
The methodology presented in this work may be used 
    to further explore candidate materials for energy storage.

\begin{acknowledgments}
We acknowledge the support of the Natural Sciences and Engineering Research Council of Canada (NSERC), [funding reference numbers RGPIN-2019-07149 and DGECR-2019-00008], as well as support from Université du Québec à Trois-Rivières. The computational resources were provided by Calcul Québec and the Digital Research Alliance of Canada.
\end{acknowledgments}

\end{document}